\documentclass[reprint,amsmath,amssymb,aps,prb,superscriptaddress]{revtex4-1}

\pdfoutput=1

\usepackage{hyperref,url}
\usepackage{amsmath}
\usepackage{amsfonts}
\usepackage[capitalise]{cleveref}
\usepackage{placeins}
\hypersetup{breaklinks,colorlinks=true,linkcolor=blue,citecolor=blue,urlcolor=blue,filecolor=blue}
\usepackage{graphicx}
\usepackage{dcolumn}

\newcommand\mat\mathbf

\newcommand{\insertrev}[1]{{\textcolor{black} {#1}}}

\begin{document}

\author{Joonho Lee}
\email{jl5653@columbia.edu}
\affiliation{Department of Chemistry, Columbia University, New York, NY 10027, USA.}
\author{Fionn D. Malone}
\email{malone14@llnl.gov}
\affiliation{Quantum Simulations Group, Lawrence Livermore National Laboratory, 7000 East Avenue, Livermore, CA 94551 USA.}
\author{David R. Reichman}
\email{drr2103@columbia.edu}
\affiliation{Department of Chemistry, Columbia University, New York, NY 10027, USA.}

\title{Note: The performance of phaseless auxiliary-field quantum Monte Carlo on the ground state electronic energy of benzene
}
\maketitle

Benchmark datasets are crucial
for developing and assessing methods for the treatment of electron correlation.
In the past several years, the Simons Collaboration on the Many-Electron Problem has produced
multiple important
benchmark studies
on the two-dimensional Hubbard model\cite{leblanc2015solutions,zheng2017stripe,qin2020absence},
transition metal molecules\cite{williams2020direct},
and hydrogen chains\cite{motta2017towards,motta2019ground}.
These studies have offered new insights
into the underlying difficulties associated with distinct electronic structure approaches in a variety of settings, 
and have provided the community
with state-of-the-art benchmarks where exact, unbiased calculations are generally not yet feasible.

The recent blind test\cite{eriksen2020benzene} by Eriksen {\em et al.} contributes to the body of knowledge in the field of computational electronic structure theory in a similar manner.
The target application of the study is the calculation of the non-relativistic Born-Oppenheimer frozen-core correlation energy of a benzene molecule in the cc-pVDZ basis set\cite{eriksen2020benzene}, with a resulting correlation space of 30 electrons and 108 orbitals. Unlike the Simons Collaboration on the Many-Electron Problem benchmark studies, the work of Ref.~\citenum{eriksen2020benzene} is focused on a single point calculation, but is completely blind, such that the authors have contributed their final results without knowledge of the exact answer or the results from other contributors. This latter aspect significantly enhances the unbiased assessment of competing and complimentary approaches.

The blind test reports the frozen-core correlation energies from a total of eight methods, all developed by the authors of Ref. \citenum{eriksen2020benzene}.
These methods can be largely grouped into five categories:
(1) one based on a many-body expansion approach (MBE-FCI),
(2) three based on a selected configuration interaction approach with a second-order perturbative correction (ASCI, iCI, and SHCI),
(3) one based on a selected coupled-cluster theory approach with a second-order perturbative correction (FCCR or more precisely FCCR(2)),
(4) one based on a matrix product state parametrization (DMRG),
and
(5) two based on the full configuration interaction quantum Monte Carlo (AS-FCIQMC and CAD-FCIQMC). 
Interested readers are referred to Ref.~\citenum{eriksen2020benzene} for further information of each method.

In the present note, we examine the accuracy of phaseless auxiliary-field quantum Monte Carlo (ph-AFQMC) on the identical problem posed by Eriksen {\em et al.}
ph-AFQMC is a method that has prominently featured in several benchmark studies led by the Simons Collaboration \cite{leblanc2015solutions,zheng2017stripe,qin2020absence,williams2020direct,motta2017towards,motta2019ground} and
has stood out as a flexible and state-of-the-art {\it ab-initio} approach\cite{zhang2003quantum, Al-Saidi2006, suewattana2007phaseless, Purwanto2008,
Purwanto2015,
motta_back_prop,
hao2018accurate,
liu2018ab,
shee2018gpu,
motta_forces,
zhang_nio,
shee2019achieving,
Shee2019,
motta_kpoint,
malone_isdf,
lee_2019_UEG,
lee2020stochastic,
lee2020utilizing,
liu2020unveiling,
malone2020gpu}.
ph-AFQMC is a projector MC method that naturally parametrizes the wavefunction in a non-linear fashion. We refer interested readers to the recent review by Motta and Zhang for details of the approach\cite{Motta2019}.
The only uncontrolled bias introduced in ph-AFQMC is the error due to the phaseless constraint\cite{zhang2003quantum} imposed via a predefined trial wavefunction.
While one must be cognizant of this bias, imposing the phaseless constraint is necessary to remove the fermionic sign (or phase) problem entirely. In other words, due to the constraint, statistical errors do not grow exponentially with system size and one does not need exponentially many walkers to cull reasonable statistics. Furthermore, as long as the underlying phaseless constraint is imposed with a size-consistent trial wavefunction, the approach guarantees size-consistency overall\cite{lee_2019_UEG}. One major potential drawback of this method is that the resulting ph-AFQMC energy is not variational\cite{carlson1999issues}.
It has been shown that ph-AFQMC can be exceptionally accurate for systems with mainly dynamic correlation (like benzene)\cite{hao2018accurate,shee2019achieving, Shee2019, lee_2019_UEG,lee2020utilizing}.
We also note that ph-AFQMC has been shown to perform well on systems with strong correlation as well, although
complicated trial wavefunctions have often been necessary in such cases\cite{zheng2017stripe,motta2017towards,motta2019ground,Shee2019,shee2019achieving,lee2020utilizing,qin2020absence}.
Given the features of ph-AFQMC outlined above, as well as the fact that it falls in a class distinct from the five categories of the examined methods, we believe that the addition of ph-AFQMC results to those of Ref.~\citenum{eriksen2020benzene} would be quite useful. We provide these results here, along with some additional observations associated with the use and accuracy of ph-AFQMC in its most scalable form.  Obviously our results are not ``blind" in the manner of those presented in Ref.~\citenum{eriksen2020benzene}, however we strive to present completely unbiased and unadjusted results which reflect standard practice and complete convergence.

The choice of trial wavefunction wholly determines the accuracy of ph-AFQMC. It is possible to exploit multi-determinant (MD) trial wavefunctions and observe a convergence of the ph-AFQMC energy with respect to the number of determinants \cite{landinez2019non,lee_2019_UEG}. While large MD trial wavefunctions may yield near-exact energies, such wavefunctions are not scalable in general and destroy the size-consistency of ph-AFQMC when the determinantal expansion is aggressively truncated.
A size-consistent, scalable trial wavefunction is a single determinant (SD) wavefunction.
The combination of ph-AFQMC with an SD trial wavefunction (e.g. Hartree-Fock (HF) orbitals\cite{lee_2019_UEG}, Kohn-Sham density functional theory orbitals\cite{Shee2019}, and approximate Brueckner orbitals\cite{lee2020utilizing}) has previously demonstrated relatively high accuracy and scalability.  In particular, in our experience the use of a HF trial with ph-AFQMC provides accuracy at roughly the CCSD(T) level while enabling the treatment of larger systems \cite{Purwanto2008,lee_2019_UEG,lee2020utilizing,malone2020gpu}. \insertrev{Further studies are necessary to understand the scope of ph-AFQMC with an SD trial relative to that of CCSD(T).}
The ph-AFQMC algorithm with an SD trial consists of the use of a quartic-scaling molecular orbital transformation of Cholesky decomposed integrals ($\mathcal O(N^4)$) once at the beginning of the QMC run, cubic-scaling propagation ($\mathcal O(N^3)$) for each time step, and the local energy evaluation ($\mathcal O(N^4)$ \insertrev{or $\mathcal O(N^3)$ with recent advances\cite{malone_isdf,motta2019efficient,lee2020stochastic}}) for each sampled MC block. \insertrev{The number of samples required for a fixed statistical error scales as $\mathcal O(N^2)$, but this overhead is practically not difficult to address due to its highly parallelizable nature.} Therefore, the method is overall \insertrev{quintic}-scaling \insertrev{with recent developments\cite{malone_isdf,motta2019efficient,lee2020stochastic}} \insertrev{for a fixed statistical error}, which is more scalable than many other state-of-the-art approaches.

Respecting the unbiased nature of this benchmark, instead of focusing on removing the phaseless error via large MD trial wavefunctions, 
we first employed the most scalable (and yet least accurate) trial wavefunction, an SD trial based on spin-restricted HF (RHF) orbitals. We refer to the resulting method as ph-AFQMC+RHF.
We emphasize that ph-AFQMC with an SD trial wavefunction is what most practitioners of ph-AFQMC would employ for general large-scale {\it ab-initio} applications.
Furthermore, we examined the improvement that one gains by using a simple MD trial wavefunction based on a complete active space self-consistent field (CASSCF) wavefunction with a 6-electrons and 6-orbitals active space ($\pi$ and $\pi^*$ orbitals). 
We refer to this method as ph-AFQMC+CAS(6,6)\cite{cas:details}.
We used QMCPACK\cite{qmcpack,qmcpack2} to run ph-AFQMC calculations on benzene. 
ph-AFQMC+RHF was run with the cc-pVDZ, cc-pVTZ, and cc-pVQZ basis sets\cite{Dunning1989} whereas ph-AFQMC+CAS(6,6) was performed only for the cc-pVDZ basis set. 
We used 1024 walkers for ph-AFQMC+RHF and 1280 walkers for ph-AFQMC+CAS(6,6). A time step of 0.005 a.u. was used. The population control bias and the time step error were found to be negligible for the purpose of this study.
Molecular integrals for QMCPACK were generated by PySCF\cite{PYSCF}. CCSD and CCSD(T) calculations were performed with Q-Chem \cite{Shao2015}. 
The smallest basis set, cc-pVDZ, was used in the blind test\cite{eriksen2020benzene}, but
we further report larger basis set results along with the T-Q extrapolated complete basis-set (CBS) energy according to Helgaker's formula\cite{helgaker1997basis}. The assessment of different approaches in the complete basis set limit is also important for the detailed evaluation of various methodologies.
\begin{table}[h!]
  \centering
  \begin{tabular}{c|c}\hline
Method & $E_\text{corr}$ (m$E_h$)\\ \hline
CCSD(T) & -859.5\\\hline
CCSDT & -859.9\\
ASCI & -860.0(2)\\
iCIPT2 & -861.1(5)\\
CCSDTQ & -862.4\\
DMRG & -862.8(7)\\
FCCR(2) & -863.0\\
MBE-FCI & -863.0\\
CAD-FCIQMC & -863.4\\
AS-FCIQMC & -863.7(3)\\
SHCI & -864(2)\\\hline
ph-AFQMC+CAS(6,6) & -864.3(4)\\
ph-AFQMC+RHF & -866.1(3)\\
\hline
  \end{tabular}
  \caption{
  The frozen-core correlation energy of benzene in the cc-pVDZ basis set
  using various methods.
  All energies other than CCSD(T) and ph-AFQMC were taken
  from the blind test results in Ref. \citenum{eriksen2020benzene}.
  \insertrev{Note that the numbers in parentheses, except those for AS-FCIQMC and ph-AFQMC, represent author-assessed uncertainties associated with a method-specific extrapolation procedure. These uncertainties are not directly comparable between different methods since they were estimated by different means as explained in the supplementary materials of Ref. \citenum{eriksen2020benzene}.}
  }
  \label{tab1}
\end{table}

As mentioned previously, ph-AFQMC+RHF places emphasis on scalability over accuracy. 
As a result of this, in \cref{tab1}, we see that ph-AFQMC+RHF deviates from the value where many methods agreed on (i.e., -863 m$E_h$) by -3.1(3) m$E_h$. We note that this deviation is comparable in magnitude to that of ASCI (+3.0(2) m$E_h$) as well as CCSDT/CCSD(T), although the direction of the deviation clearly reflects the non-variational nature of ph-AFQMC.
This ph-AFQMC+RHF calculation required modest computation resources: 4 hours with 32 graphics processing units (GPUs) (NVIDIA V100(VOLTA); 4 GPUs per node; 16GB per GPU). On the other hand, ph-AFQMC+CAS(6,6) is more accurate than ph-AFQMC+RHF while its scalability is ultimately limited by the CAS calculation itself for general applications. The resulting ph-AFQMC+CAS(6,6) energy deviation from the ``exact" answer (-863 m$E_h$) is -1.3(4) $mE_h$, which is about a factor of 2.4 improvement over ph-AFQMC+RHF. \insertrev{This deviation is comparable to that of SHCI (-1(2) $mE_h$) and highlights the accuracy of ph-AFQMC with a relatively simple MD trial wavefunction. It should be noted that more accurate ASCI and SHCI post blind test results may be found in the supplementary materials of Ref. \citenum{eriksen2020benzene}.}
The ph-AFQMC+CAS(6,6) calculation was performed on 160 cores (Intel(R) Xeon(R) Gold 6148 CPU @ 2.40GHz; 40 cores per node) for 10 hours.

\begin{table}[h!]
  \centering
  \begin{tabular}{c|c|c|c}\hline
basis & ph-AFQMC+RHF & CCSD & CCSD(T)\\ \hline
cc-pVTZ & -1033.7(3)  &-975.2 & -1027.3\\\hline
cc-pVQZ & -1085.5(4) &-1027.3 & -1079.0 \\ \hline
CBS & -1123.3(7) & -1057.4 &-1116.7\\ \hline
  \end{tabular}
  \caption{
  The frozen-core correlation energy (m$E_h$) of benzene
  using ph-AFQMC+RHF, CCSD, and CCSD(T) in the cc-pVTZ and cc-pVQZ basis sets.
  }
  \label{tab2}
\end{table}
Lastly, we report the larger basis set ph-AFQMC+RHF results and its CBS limit energy in \cref{tab2}.
The correlation space increases from 108 orbitals (cc-pVDZ) to 258 orbitals (cc-pVTZ) and 504 orbitals (cc-pVQZ).
Similarly to cc-pVDZ, ph-AFQMC+RHF correlation energies are 6-7 m$E_h$ lower than those of CCSD(T) in both bases and in the complete basis set limit.
We expect converged ph-AFQMC with an MD trial wavefunction to lie between these two numbers similarly to cc-pVDZ. The same computational resource as that of ph-AFQMC+RHF/cc-pVDZ was used for cc-pVTZ and 64 GPUs were used for 4 hours for cc-pVQZ.

In summary, we report ph-AFQMC correlation energies for the problem posed in the recent blind test of Ref.~\citenum{eriksen2020benzene}, namely, a benzene molecule in the cc-pVDZ basis set. In addition, we report the ph-AFQMC+RHF correlation energies on larger basis sets (cc-pVTZ and cc-pVQZ) along with the extrapolated complete basis set limit correlation energy. We believe that due to the accuracy, flexibility and scalability of the approach, the addition of ph-AFQMC results to those of the recent blind test will contribute to the informed use of a broad set of methods to tackle diverse electronic structure problems. Challenges for objective benchmark studies include the broad coverage of relevant methods, the choice of representative targets \insertrev{such as energy differences}, and the study of systems as close to the complete basis set limit as possible. The recent studies by the Simons Many-Electron Collaboration\cite{leblanc2015solutions,zheng2017stripe,qin2020absence,williams2020direct,motta2017towards,motta2019ground} and by Eriksen and {\em et al.}\cite{eriksen2020benzene} illustrate community efforts towards this goal, to which we hereby add restricted but useful information concerning the ph-AFQMC approach.

{\it Data availability} The data that supports the findings of this study are available within the article.

{\it Acknowledgement} D.R.R. was supported by Grant No. NSF-CHE 1954791. The work of FDM was supported by the U.S. Department of Energy (DOE), Office
of Science, Basic Energy Sciences, Materials Sciences and Engineering Division, as part of the Computational Materials Sciences
Program and Center for Predictive Simulation of Functional Materials (CPSFM). 
The work of FDM was performed under the auspices of the U.S. DOE by LLNL under Contract No. DE-AC52-07NA27344. Some of AFQMC calculations received computing support from the LLNL Institutional Computing Grand Challenge program.

\bibliography{main}

\begin{thebibliography}{38}
\bibitem [{cas()}]{cas:details}%
  \BibitemOpen
  \href@noop {} {}\bibinfo {note} {The determinantal expansion in the CAS(6,6)
  trial wavefunction was truncated by a coefficient threshold of 0.999999,
  which yielded a variational energy that is essentially identical to the full
  determinantal expansion. The resulting truncated trial wavefunction consists
  of a total of 87 determinants. We further note that CASSCF changes the 1s
  core orbitals of carbon atoms from those of RHF, but the corresponding core
  relaxation lowers the CASSCF energy only by 0.8 $\mu$$E_h$. Such a small
  effect is negligible compared to the statistical error in ph-AFQMC and
  therefore we ignore this.}\BibitemShut {Stop}%
\end{thebibliography}

\end{document}